\documentclass[letterpaper%
,pra%
,twocolumn%
,notitlepage%
,superscriptaddress%
,nofootinbib%
]{revtex4-1}

\PassOptionsToPackage{colorlinks}{hyperref}

\usepackage {amsmath}
\usepackage{graphicx}
\usepackage{braket}
\usepackage{amsthm}
\usepackage{amssymb}
\usepackage{url}
\usepackage{color}
\usepackage{mathtools}
\usepackage{enumitem}
\usepackage{soul}

\usepackage{bm}

\usepackage{tikz}
\usetikzlibrary{decorations}
\usetikzlibrary{patterns,snakes}

\newtheorem{theorem}{Theorem}

\makeatletter
\@ifundefined{textcolor}{}
{%
 \definecolor{BLACK}{gray}{0}
 \definecolor{WHITE}{gray}{1}
 \definecolor{RED}{rgb}{.9,0,0}
 \definecolor{GREEN}{rgb}{0,.6,0}
 \definecolor{BLUE}{rgb}{0,0,.9}
 \definecolor{CYAN}{cmyk}{1,0,0,0}
 \definecolor{MAGENTA}{cmyk}{0,1,0,.2}
 \definecolor{YELLOW}{cmyk}{0,0,1,0}
 }
\makeatother

\providecommand{\abs}[1]{\lvert#1\rvert}

\newcommand{\reals}{\mathbb{R}}

\newcommand\tp{\mathrm{T}}

\DeclareMathOperator{\tr}{tr}

\DeclareMathOperator{\sech}{sech}

\renewcommand{\Im}{\operatorname{Im}}

\def\group#1{\mathrm{#1}}

\def\op#1{\hat{#1}}
\def\opvec#1{\op{\vec{#1}}}

\def\id{I}

\def\1{\mat{\id}}

\def\mat#1{\bm{\mathrm{#1}}}

\renewcommand{\vec}[1]{\bm{\mathrm{#1}}}

\def\abs#1{\left\lvert{#1}\right\rvert}


\usepackage{hyperref}

\graphicspath{ {./Figures/} }

\allowdisplaybreaks

\begin{document}

\title{Universal quantum computation with temporal-mode bilayer square lattices}
\author{Rafael N. Alexander} \email[Email: ]{rafaelalexander@unm.edu} \affiliation{Center for Quantum Information and Control, University of New Mexico, MSC07-4220, Albuquerque, New Mexico 87131-0001, USA}
\affiliation{Department of Physics,  University of Virginia, Charlottesville, Virginia 22903, USA}\affiliation{Centre for Quantum Computation and Communication Technology, School of Science, RMIT University, Melbourne, Victoria 3001, Australia} 
\author{Shota Yokoyama}\affiliation{Centre of Quantum Computation and Communication Technology, School of Engineering and Communication Technology, University of New South Wales, Canberra, Australian Capital Territory 2600, Australia}
 \author{Akira Furusawa}\affiliation{Department of Applied Physics, School of Engineering, The University of Tokyo, 7-3-1 Hongo, Bunkyo-ku, Tokyo 113-8656, Japan}
 \author{Nicolas C. Menicucci} \email[Email: ]{ncmenicucci@gmail.com}\affiliation{Centre for Quantum Computation and Communication Technology, School of Science, RMIT University, Melbourne, Victoria 3001, Australia} 

\date{\today}
\begin{abstract}
We propose an experimental design for universal continuous-variable quantum computation that incorporates recent innovations in linear-optics-based continuous-variable cluster state generation and cubic-phase gate teleportation. The first ingredient is a protocol for generating the bilayer-square-lattice cluster state (a universal resource state) with temporal modes of light. With this state, measurement-based implementation of Gaussian unitary gates requires only homodyne detection. Second, we describe a measurement device that implements an adaptive cubic-phase gate, up to a random phase-space displacement. It requires a two-step sequence of homodyne measurements and consumes a (non-Gaussian) cubic-phase state. 
\end{abstract}
\maketitle
 
\section{Introduction}\label{sec:introduction}

Recently, there has been substantial progress in developing the basic building blocks of quantum computation using continuous-variable cluster states (CVCSs)~\cite{Menicucci2006}. Two-dimensional (or higher-dimensional) cluster states enable universal quantum computation provided that one can also perform suitable sequences of single-site measurements~\cite{Raussendorf2001, Menicucci2006}. Thus, much effort has focused on generating large-scale cluster states and implementing sets of measurements that enable universal measurement-based quantum computation (MBQC)~\cite{Gu2009}.  

The scalability of one-dimensional CVCS architectures is now well established. Recent experimental demonstrations have yielded states consisting of 60 frequency modes (where each mode is simultaneously addressable)~\cite{Chen2014} and greater than one million temporal modes (where each mode is sequentially addressable) with no significant experimental obstacles hindering the generation of even larger states~\cite{Yokoyama2013, Yoshikawa2016}. These states are endowed with a multi-layered graph~\footnote{Here \emph{graph} has a precise mathematical meaning, being uniquely defined for all Gaussian pure states up to phase-space displacements and overall phase as in Ref.~\cite{Menicucci2011}. }, from which can be derived a simple and compact state generation circuit consisting of offline squeezers and constant-depth linear optics~\cite{Menicucci2011a}. This feature makes them highly compatible with multi-mode squeezing platforms, such as optical parametric oscillators (OPOs)~\cite{Menicucci2007, Menicucci2008, Flammia2009, Menicucci2011a} and Josephson traveling-wave parametric amplifiers~\cite{Grimsmo2017}.
Though these states generalize to two~\cite{Alexander2016b, Menicucci2011a, Wang2014} and higher dimensions~\cite{Wang2014}, the generation of such a scalable universal cluster state has yet to be demonstrated. 

Complementary to this work is an effort to implement a universal set of continuous-variable gates using tailored small-scale resource states (prepared offline) and gate teleportation~\cite{Lloyd1999}. Examples include demonstrations of single-mode Gaussian operations (including linear optics and squeezing)~\cite{Filip2005, Yoshikawa2007,  Yoshikawa2011}, single-mode non-Gaussian operations such as the cubic-phase gate~\cite{Yukawa2013, Yukawa2013a, Marshall2015a, Miyata2016, Arzani2017}, and the two-mode sum gate~\cite{Yoshikawa2008}. Such demonstrations are gate-specific and have remained limited to small-scale implementations. It is therefore desirable to synthesize these elements into one universal, scalable design. 

Here we consider an approach to universal quantum computation that seeks to marry the above research threads. We begin by modifying an existing protocol (proposed by some of us) for generating a universal cluster state known as the \emph{bilayer square lattice}~\cite{Alexander2016b} (BSL). Our implementation makes use of temporal modes, and requires four squeezers, five beamsplitters, and two delay loops. The practicality of this scheme stems from its similarity to the aforementioned one-dimensional experiment~\cite{Yoshikawa2016}. 

Beyond state generation, much of the previous work on quantum computing with CVCSs has focused on Gaussian operations~\cite{Ukai2010, Ukai2011, Su2013, Miyata2014, Ferrini2013a, Alexander2014, Yokoyama2015, Ferrini2016, Alexander2016}, which can be implemented with non-adaptive homodyne measurements~\cite{Menicucci2006, Gu2009} and can be efficiently simulated~\cite{Bartlett2002}. Universal quantum computation requires non-Gaussian gates, which can be implemented by measuring quadratic or higher order polynomials in the quadrature operators or by injecting  non-Gaussian single-mode states directly into the cluster state~\cite{Gu2009}. In order for non-Gaussian computations to proceed deterministically, the measurements must be adaptive~\cite{Gu2009}. 

Other than the present proposal, Ref.~\cite{Gu2009} is the only explicit experimental proposal for implementing a universal set of measurement-based gates on a CVCS. In that approach, a non-Gaussian gate known as the \emph{cubic-phase gate} is implemented on a CVCS using  photon-number-resolving detection and adaptive squeezing operations. Even in the absence of experimental sources of error (such as lost photons), requiring adaptive squeezing operations is problematic. This is because the amount of squeezing required is unbounded (it depends on the random homodyne measurement outcomes), and implementing high levels of squeezing via MBQC is known to amplify the intrinsic noise due to the Gaussian nature of the cluster state~\cite{Alexander2014}.

 Here we propose a measurement device that consumes a cubic-phase ancilla state and implements a cubic-phase gate. The measurement adaptivity that compensates for the randomness of the measurement outcomes can be implemented with a single adaptive phase shift element, and therefore the required amount of squeezing in this protocol is not probabilistic. Our device uses a two-step adaptive homodyne measurement, similar to the gate teleportation scheme proposed in Ref.~\cite{Miyata2016}.
 
This paper is organized as follows. In Section~\ref{sec:defs}, we introduce notation. In Section~\ref{sec:stategen}, we present our experimental design for generating the bilayer square lattice cluster state using temporal modes. Using the cluster state's graph, we construct  a convenient set of observables for experimental methods of entanglement verification. In Section~\ref{sec:mbgates}, we demonstrate how universal quantum computation proceeds via measurement of the resource state. We conclude in Section~\ref{sec:conc}.

\section{Notation and definitions}\label{sec:defs}
%
Throughout this paper, we adopt the same conventions as in Ref.~\cite{Alexander2016b}. For convenience, we summarize them here. For all modes: 
\begin{align}
\op q \coloneqq& \tfrac {1} {\sqrt 2} (\op a + \op a^\dag), \\
\op p \coloneqq& \tfrac {1} {i\sqrt 2} (\op a - \op a^\dag).
\end{align}
Using ${[\op a, \op a^\dag] = 1}$, this implies that ${[\op q, \op p] = i}$ with ${\hbar = 1}$ and vacuum variances $\langle \hat{q}^{2} \rangle_{\text{vac}}=\langle \hat{p}^{2} \rangle_{\text{vac}}=\frac{1}{2}$. Define the rotated quadrature operators:
\begin{align} 
\hat{q}(\theta) \coloneqq \hat{q}\cos \theta  - \hat{p}\sin \theta, \\
\hat{p}(\theta) \coloneqq \hat{p}\cos \theta   + \hat{q}\sin \theta.
\end{align}
 We use the following column notation for operators on $n$ modes:
 \begin{align}
 \hat{\mathbf{q}}\coloneqq & (\hat{q}_{1}, \dots \hat{q}_{n})^\tp, \\
  \hat{\mathbf{p}}\coloneqq &(\hat{p}_{1}, \dots \hat{p}_{n})^\tp, \\ 
 \hat{\mathbf{x}}\coloneqq & \begin{pmatrix} \hat{\mathbf{q}} \\ \hat{\mathbf{p}}\end{pmatrix}.
 \end{align}
Columns of $c$-numbers  corresponding to these operators are denoted analogously as $\mathbf{q}$, $\mathbf{p}$, and $\mathbf{x}$.
Let $\ket{s}_{q_{i}}$ and $\ket{t}_{p_{i}}$ $\forall s, t \in \mathbb{R}$ denote eigenstates of position and momentum for mode $i$, respectively. 
In this proposal, we show how to implement the universal gate set shown in Table~\ref{tab:unigates}.
\begin{table}
 \begin{tabular}{|c | c | c | c|}  \hline 
Gate & Class & Symbol & Equation \\
  \hline \hline
$\hat{q}$-shift & I  & $\hat{X}(s)$ & $\exp(-i s \hat{p}) $\\
 \hline
 $\hat{p}$-shift & I  & $\hat{Z}(t)$ & $\exp(i t \hat{q})$ \\
 \hline
phase delay & II & $\hat{R}(\theta)$ & 
$ \exp(i\theta\hat a^\dag \hat a)$ \\
 \hline
squeezing & II & $\hat{S}(r)$ & 
$\exp[-r(\hat a^2 - \hat a^{\dag 2})/2]$ \\
 \hline
shear & II & $\hat{P}(\sigma)$ & $\exp(i \sigma \hat{q}^{2}/2)$\\
 \hline
controlled-Z & II & $ \hat{C}_{Z}(g)$ & $\exp(i g \hat{q}\otimes\hat{q})$\\
 \hline
cubic-phase & III & $ \hat{K}(\chi)$ & $  \exp(i \chi\hat{q}^{3}/3 ) $ \\
 \hline
\end{tabular}
\caption{A universal set of gates for continuous-variable quantum computation. Here, the parameters ${s, t, r, \sigma, g, \chi \in \mathbb{R}}$ and ${\theta\in [0, 2\pi)}$. Gates in class I generate all phase-space displacements. Gates in the union of classes I and II (redundantly) generate the Gaussian unitary group. Class III contains the cubic-phase gate, which extends the Gaussian unitaries to a universal gate set~\cite{Lloyd1999}. Note that the squeezing gate is sometimes defined with respect to the \emph{squeezing factor} $s=e^{r}$, which is the rescaling factor for the Heisenberg picture evolution of the quadrature operator $\hat{q}$ under the squeezing operation~\cite{Alexander2016b}. In this paper we reserve the letter $s$ for displacements in position and work only with the squeezing parameter $r$. }\label{tab:unigates}
\end{table}

Below, we define some additional gates and states required to generate the BSL.
We define the two-mode beam-splitter gate
\begin{align}
	\hat{B}_{ij}(\theta)\coloneqq & \exp\left[-\theta \left(\op a_i^\dag \op a_j - \op a_j^\dag \op a_i\right) \right] \nonumber \\
	= & \exp\left[-i\theta\left(\hat{q}_{i}\hat{p}_{j}-\hat{q}_{j}\hat{p}_{i}\right)\right],
\end{align}
where $\sin \theta$ is the reflectivity of the beam splitter. 
Its Heisenberg action on $\op{\mathbf{x}}=(\op{q}_{i}, \op{q}_{j}, \op{p}_{i}, \op{p}_{j})^\tp$ is given by
\begin{align}
	\mathbf{B}_{ij}(\theta) = \begin{pmatrix} \cos{\theta} & -\sin{\theta} & 0 & 0 \\ \sin{\theta} & \cos{\theta} &  0 & 0 \\ 0 & 0 & \cos{\theta} & -\sin{\theta} \\ 0 & 0 & \sin{\theta} & \cos{\theta} \end{pmatrix}.
\end{align}
For the special case of a 50:50 beamsplitter, which corresponds to $\theta=\frac{\pi}{4}$, we drop explicit angle dependence:
\begin{align}
\hat{B}_{ij} \coloneqq \hat{B}_{ij}\left(\frac{\pi}{4}\right).
\end{align}
We define the single-mode squeezed states
\begin{align}
\ket{\eta(r)}\coloneqq \hat{S} (r)\ket{0}
\end{align}
where $\ket{0}$ is the vacuum state and $r>0$ ($r<0$) corresponds to squeezing in the momentum (position) quadrature. We define the idealized (and unnormalizable) cubic-phase states
 \begin{align}
 \ket{\phi_{\chi}}\coloneqq \int \mathrm{d}s \, e^{i \chi s^{3}/3} \ket{s}_{q}. \label{eq:cubicstate}
 \end{align}
Approximations to the cubic-phase states can be created as a truncated superposition of the Fock states ${\ket{\phi_{\chi}} = \sum_n c_n \ket{n}}$ with  two-mode-squeezed vacuum states and photon detection~\cite{Yukawa2013}. Once prepared, such states can be stored with cascaded cavities for on-demand injection into the BSL~\cite{Yoshikawa2013}. 

\subsection{Graphical Notation} 
Below, we use the graphical calculus for Gaussian pure states~\cite{Menicucci2011} in order to describe the generation procedure for the BSL and the implementation of gates. Gaussian pure states with zero mean are one to one (up to an overall phase) with complex, symmetric adjacency matrices (i.e., $\ket{\psi} \leftrightarrow \mathbf{Z}$)~\cite{Menicucci2011}. This is immediate from the position-space representation
\begin{align}
\psi(\mathbf{q}) =\braket{\mathbf{q}\vert\psi} = \frac{( \det \Im \mat Z)^{1/4}}{\pi^{n/4}}\exp\left[\frac{i}{2}\mathbf{q}^\tp\mathbf{Z}\mathbf{q}\right] \label{eq:graphpsi}
\end{align}
where $\ket{\mathbf{q}} \coloneqq \bigotimes_{i=1}^{n} \ket{q_{i}}_{q}$. 
Note that $\Im\mathbf{Z}$ is required to be positive definite  in order for the state to be normalizable. An independent set of $n$ nullifiers for $\ket{\psi}$ is given by:
\begin{align}
	\left(\hat{\mathbf{p}} - \mathbf{Z}\hat{\mathbf{q}}\right)\ket{\psi}= \mathbf{0}. \label{eq:standardnullifier}
\end{align}
This set forms a basis for the commutative algebra of linear nullifiers of $\ket{\psi}$.

For convenience, when using the graphical calculus in figures and diagrams, we restrict ourselves to the conventions of the \emph{simplified graphical calculus} introduced in Refs.~\cite{Menicucci2011a, Alexander2016b}. Therefore, self-loops (corresponding to the diagonal entries of $\mathbf{Z}$) are not drawn but are assumed to have weight $i\sech{2r}$, where $r$ parametrizes the amount of vacuum squeezing used to generate the state.  Furthermore, edges between distinct nodes have the same magnitude $\mathcal{C} \tanh{2r}$, where $\mathcal{C}$ is indicated below each graph. These edges have a phase of $\pm 1$, denoted by blue and yellow coloring, respectively. 

\section{Temporal-mode generation of the bilayer square lattice}\label{sec:stategen}

\subsection{Construction}
\begin{figure*}
\includegraphics[width=1\linewidth]{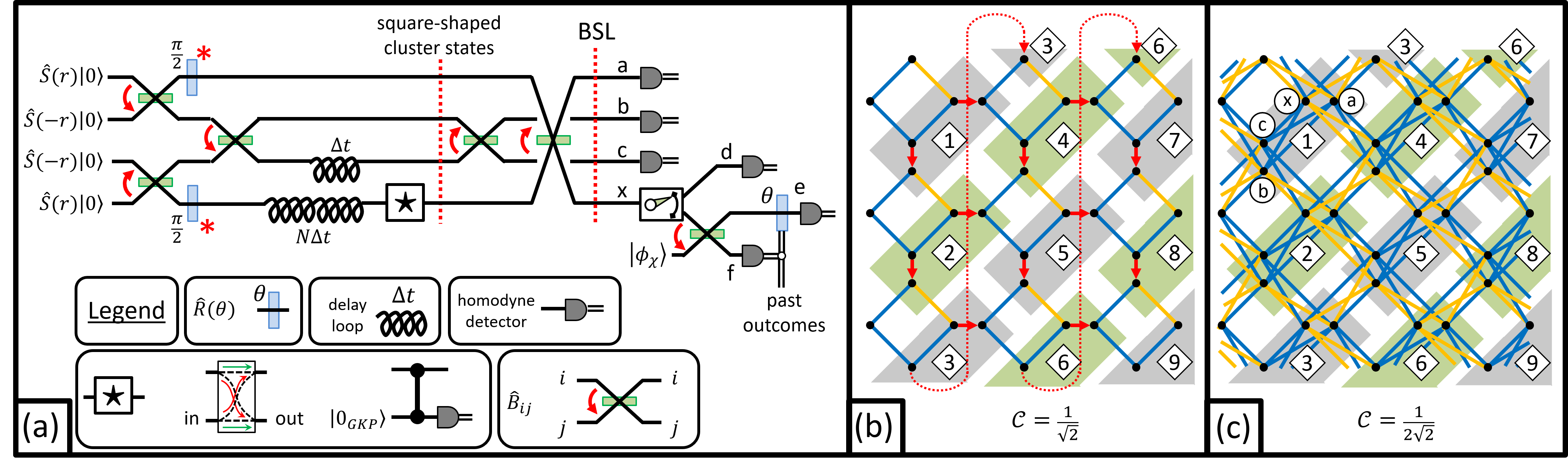}
\caption{$\mathbf{(a)}$ Optical circuit diagram that executes all stages of the generation and measurement of the bilayer square lattice using temporal modes. Input states encoded within the cluster state can be addressed at the black star (``$\star$"). The legend includes two examples: (left) a switching device for encoding or removing inputs from the cluster state and (right) a syndrome-measurement circuit for quantum error correction using Gottesman-Kitaev-Preskill qubits~\cite{Menicucci2014}. The first dashed red line indicates the point at which the state is made up of disjoint squares [see panel (b)]. The second dashed red line is the point where the final cluster state exists [see panel (c)]. Cluster-state modes are measured at detectors `a', `b', `c', or `x'. Detector `x' has two settings: homodyne measurement by using `d' or cubic-phase gate-teleportation measurement by using `e' and `f'. The latter involves injecting a cubic-phase-state ancilla $\ket{\phi_{\chi}}$ and using an adaptive (variable) phase delay. See main text for details. If the `x' detector is always set to `d', then the phase delays marked by a red asterisk can be omitted by compensating with a $\pi/4$ phase delay before detectors `a', `b', `c', and `x'. $\mathbf{(b)}$ Each gray or green rectangle contains modes that arrive at the first red dotted line in panel (a) simultaneously. The numerical labels indicate the relevant time step in units of $\Delta t$. The red arrows between each disjoint square represent the application of the beamsplitters between the red dotted lines in panel (a). $\mathbf{(c)}$ The bilayer-square-lattice  graph. The letters $\{\text{a}, \text{b}, \text{c}, \text{x}\}$ in the rectangle labeled ``1" indicate at which detector each mode is measured.}\label{fig:bslconstruct}
\end{figure*}

The first step in generating the BSL with temporal modes is to create a four-mode square-shaped cluster state. The construction of these is shown graphically as follows:
\begin{align}
\includegraphics[width=0.6\linewidth]{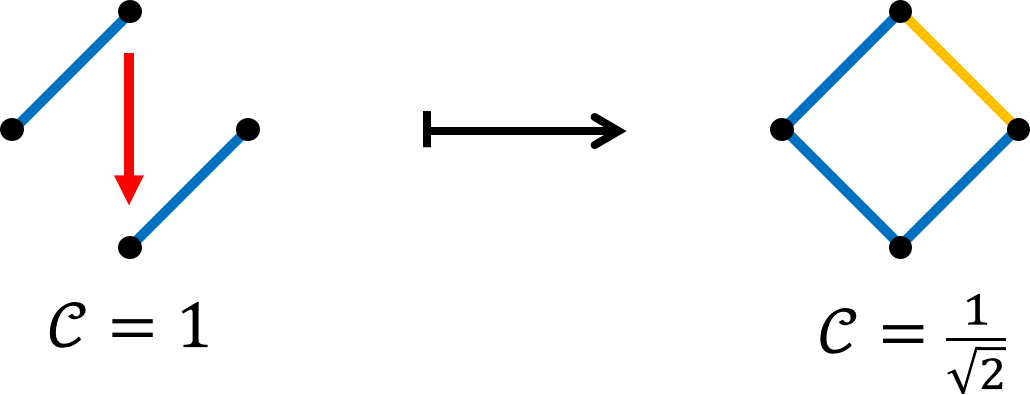}\label{eq:p2s}
\end{align} 
A pair of two-mode CVCSs [shown left]\footnote{These are equivalent to two-mode squeezed states, up to local phase delays.} are transformed by a single 50:50 beamsplitter $\hat{B}_{ij}$ [indicated by the red (downward pointing) arrow from mode $i\rightarrow j$], resulting in the square-shaped cluster state [shown to the right in Eq.~\ref{eq:p2s}].

Using temporal modes with time-bin windows of size $\Delta t$, we can generate an array of square-shaped cluster states using the optical circuit shown in Fig.~\ref{fig:bslconstruct}(a). Let $N M \Delta t$ be the runtime of the experiment, where $N$ sets the delay on the longer delay loop in Fig.~\ref{fig:bslconstruct}(a). The time axis is partitioned into $M$ segments of length $N\Delta t$, yielding an array of size $N\times M$ in the time domain. In Fig.~\ref{fig:bslconstruct}(b) we show the $N=M=3$ case as an example. 

The final step involves applying a sequence of balanced beamsplitters between modes as indicated by the red arrows in Fig.~\ref{fig:bslconstruct}(b). The resulting state has a double-layered square-lattice graph, as shown in Fig.~\ref{fig:bslconstruct}(c). The experimental simplicity of our scheme is self-evident.

\subsection{Nullifiers}

Here we discuss the nullifiers of the BSL, which play an important role in verifying genuine multipartite entanglement via homodyne detection~\cite{vanLoock2003, Walborn2009, Toscano2015, Gerke2016}. A class of experimentally convenient methods for verifying multiparite entanglement has been developed based on the van Loock-Furusawa criterion~\cite{vanLoock2003, Yokoyama2013}. Such methods can be applied to states possessing a generating set of nullifiers $\{\hat{\nu}_{i}\}_{i=1, \dots n}$ with the property that each $\hat \nu_{i}$ is a linear combination of either position or momentum quadrature operators (not both). In this case, verifying multipartite entanglement requires only two types of homodyne measurements (on distinct copies of the state): all modes in the $\hat{q}$ basis or all modes in the $\hat{p}$ basis. Variances for all nullifiers can be inferred by taking linear combinations of the data, signaling the presence of multipartite entanglement~\cite{vanLoock2003}.

The BSL graph $\mathbf{Z}_{\text{BSL}}$ in Fig.~\ref{fig:bslconstruct}(c) immediately provides a set of $n$ generators for the algebra of nullifiers [see Eq.~(\ref{eq:standardnullifier})]. These can be made Hermitian by taking the infinite-squeezing limit, $\mathbf{V}_{\text{BSL}} \coloneqq \lim_{r\rightarrow\infty} \mathbf{Z}_{\text{BSL}}$, resulting in \emph{approximate} nullifiers~\cite{Menicucci2011}:
\begin{align}
\left(\hat{\mathbf{p}} -\mathbf{V}_{\text{BSL}}\hat{\mathbf{q}}\right)\ket{\psi_{\text{BSL}}}  \to  \mathbf{0}
\end{align}
where ``$\to$" indicates the limit of infinite squeezing within $\ket{\psi_{\text{BSL}}}$ (which is independent of the limit used to define~$\mat V_{\text{BSL}}$).  As noted above, it is more convenient from an experimental viewpoint to work with nullifiers that are combinations of either position or momentum operators. Though CVCSs cannot have such a set of nullifier generators~\cite{Menicucci2011}, in some cases, we can use local phase delays (which preserve entanglement properties of the state) to transform them into states that do. This is formalized by the following theorem.  
\begin{theorem}
Any $2n$-mode, ideal (i.e., infinitely squeezed) continuous-variable (CV)  cluster state with a trace-zero, self-inverse graph  $\mat V$ can be approximated by a $2n$-mode  finitely-squeezed cluster state $\ket{\Psi_r}$, with overall squeezing parameter~$r$, that is equivalent under local phase delays to a Gaussian pure state $\ket{\Phi_r}$ that satisfies the following two properties.
\begin{enumerate}
\item $\ket{\Phi_r}$ is approximately nullified by a set of  $2n$ independent local operators that each consist of linear combinations of either position or momentum operators. Here locality is defined with respect to  $\mat V$.
\item The corresponding graph $\mathbf{Z}_{\Phi,r}$ only differs from  $\mathbf{Z}_{\Psi,r}$ by a  nonzero, uniform  reweighting of the edges and  a (different) nonzero, uniform reweighting of the  self-loops.
\end{enumerate}\label{thm:1}
\end{theorem}

\noindent For proof, see Appendix~\ref{sec:thmproof}.

The locality of these nullifiers tends to be a useful property when it comes to deriving bounds for entanglement witnesses based on the nullifier variances~\cite{vanLoock2003, Yokoyama2013}. In the  infinite-squeezing limit, the BSL graph is both self-inverse and trace zero. Therefore Theorem~\ref{thm:1} applies. 

Denote the BSL state by $\ket{\Psi_{\text{BSL}}}$. Following the abovementioned notational convention, we define the Gaussian pure state 
\begin{align}
\ket{\Phi_{\text{BSL}}}\coloneqq\hat{R} \left(\frac{\pi}{4}\right)^{\otimes 2n} \ket{\Psi_{\text{BSL}}}. \label{eq:phidef}
\end{align}
Though our proposal generates $\ket{\Psi_{\text{BSL}}}$, it could easily be modified [e.g., by including the local phase shifts in Eq.~(\ref{eq:phidef}) at the BSL line in Fig.~\ref{fig:bslconstruct}(a)] to generate $\ket{\Phi_{\text{BSL}}}$ instead. Furthermore, $\ket{\Phi_{\text{BSL}}}$ is an equivalent resource for measurement-based quantum computation and possesses a set of approximate nullifiers,
\begin{align}
	\begin{pmatrix}
		(\mat I_{2n} - \mat V_{\text{BSL}}) \opvec p
	\\
		(\mat I_{2n} + \mat V_{\text{BSL}}) \opvec q
	\end{pmatrix}
	\ket {\Phi_{\text{BSL}}} 
	 \to 
	\vec 0,
\end{align}
that are experimentally convenient for entanglement verification. 

Other known examples of cluster states with self-inverse, trace-zero graphs include the dual-rail wire~\cite{Menicucci2011a, Alexander2014, Wang2014} and the quad-rail lattice~\cite{Menicucci2011a, Alexander2016, Wang2014}.

\section{Universal Measurement-based quantum computation}\label{sec:mbgates}

Now we describe universal quantum computation with the BSL. It was shown explicitly in Ref.~\cite{Alexander2016b} how arbitrary Gaussian unitary gates can be implemented on the BSL via homodyne detection. This analysis also included the effects of finite squeezing. Here we focus on extending this scheme by adding cubic-phase state injection for implementing non-Gaussian gates. With this resource, the only type of measurement our protocol requires is homodyne detection. 

Temporal-mode architectures rely on fast control of the local-oscillator (LO) beam phase at each homodyne detector in order to set the measurement basis for each mode independently. Dynamic phase control for time bins of $~160$ ns (used in recent cluster-state demonstrations~\cite{Yokoyama2013, Yoshikawa2016}) is within the scope of current technology. \footnote{The dynamic changing of a LO beam phase has been experimentally demonstrated in Ref.~\cite{Miyata2014}. The phase control employs a waveguide phase modulator with a bandwidth of $>1$ GHz. The corresponding rise time of $<1$ ns is adequately smaller than 160ns time bins. Note that a 1 MHz operational bandwidth mentioned in Ref.~\cite{Miyata2014} is limited by other elements such as an OPO and a homodyne detector. Recently, the bandwidths of these elements have been drastically improved to $>$ 100MHz~\cite{Shiozawa2018}.}

Just like conventional square-lattice cluster states~\cite{Raussendorf2001, Menicucci2006}, the BSL can be treated as a collection of quantum wires embedded within a two-dimensional plane~\cite{Alexander2016b}. Entangling gates between these quantum wires are mediated by local projective measurements on sites that lie between wires. However, the BSL differs from other cluster states in that the quantum wires actually run along the diagonals of each square-like unit cell, rather than along the edges (which would be more conventional).

Each BSL lattice site, a.k.a.~a \emph{macronode}, consists of two modes, labeled $\alpha$ and $\beta$. Without loss of generality, assume that mode~$\alpha$ is measured at either detector~`x' or detector~`b', and similarly, assume mode~$\beta$ is measured at either~`a' or~`c'. 
We define the symmetric (+) and anti-symmetric ($-$) mode of each macronode via:
\begin{align}
\hat{a}_{\pm}\coloneqq \frac{1}{\sqrt{2}}(\hat{a}_{\alpha}\pm\hat{a}_{\beta}) = \hat{B}_{\alpha\beta} \hat{a}_{\alpha (\beta)}\hat{B}_{\alpha\beta}^{\dagger} \label{eq:modelabs}
\end{align}
This offers an alternative tensor-product decomposition of each macronode to the one provided by the physical modes, $\alpha$ and $\beta$. 
We divide the discussion of gate implementation into two parts: single-mode gates and two-mode gates. We leave an analysis of the finite-squeezing effects to future work. 

\subsection{Single-mode gates} \label{sec:smgates}

\begin{figure}
\includegraphics[width=\linewidth]{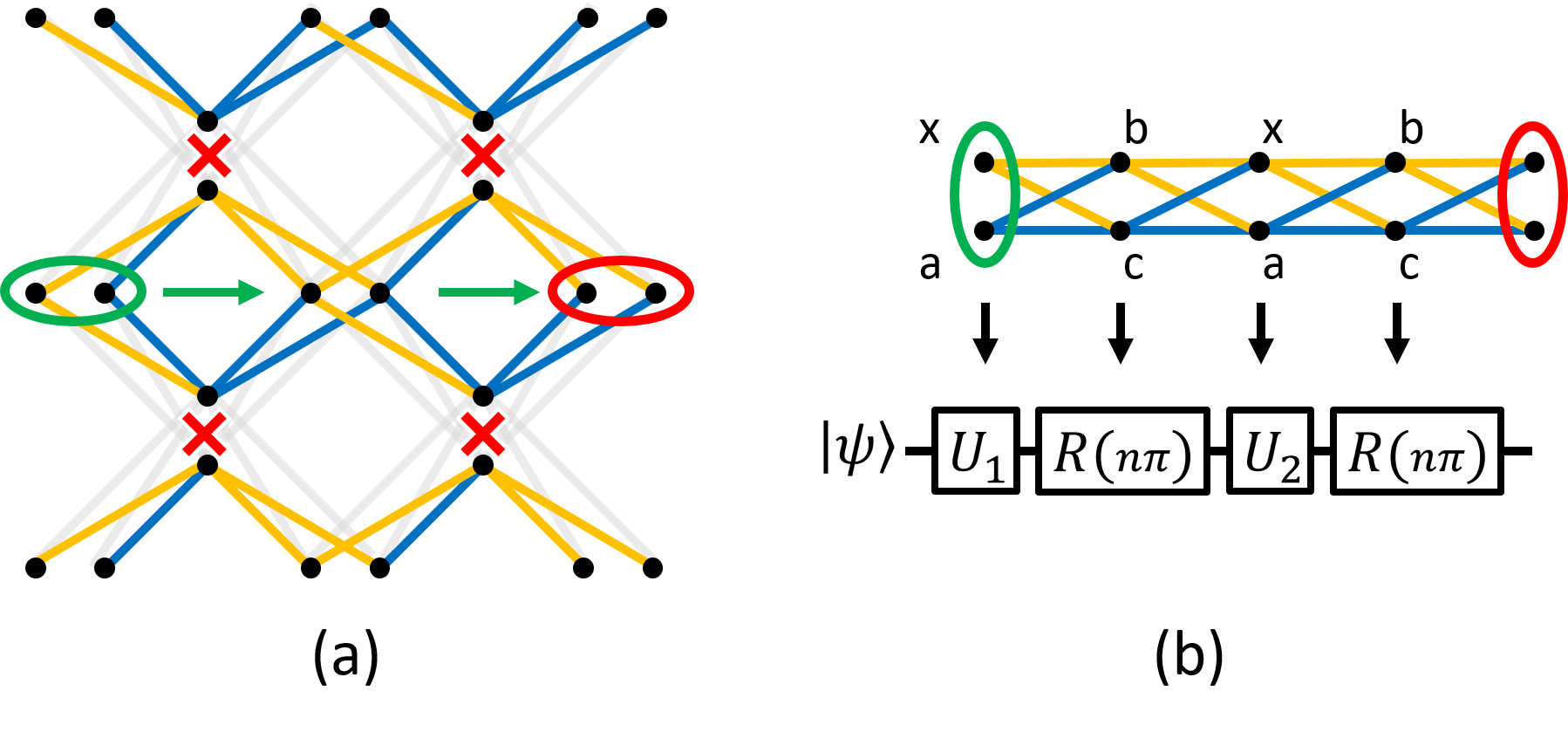}
\caption{$\mathbf{(a)}$ Measuring $\hat{q}[(-1)^{\xi} \frac{\pi}{4}]$ at detectors `b'~and `c'~(indicated by a red `X') at time index $\xi$ mod $N$ deletes the grayed-out edges~\cite{Alexander2016b}. This decouples the central three rows of nodes from the rest of the graph. An input state encoded within the green site on the left will propagate in the direction of the  green arrows. $\mathbf{(b)}$ After deletion as in panel (a), the middle three graph modes are equivalent to the wire graph shown at the top. This state is known as the dual-rail wire~\cite{Yokoyama2013, Alexander2014}.  The letters \{a, b, c, x\} indicate the detector at which the relevant nodes are measured in Fig.~\ref{fig:bslconstruct}(a). Measurements at detectors `x' and `a' can be chosen freely to implement gates, whereas measurement settings at `b' and `c' are fixed by the deletion measurements. } \label{fig:wirebsl}
\end{figure}

To implement single-mode gates, quantum wires must be decoupled from the rest of the lattice by deleting unwanted edges. This can be achieved by using detectors `b'~and `c'~to measure in the basis $\hat{q}[(-1)^{\xi} \frac{\pi}{4}]$  at particular macronodes, where $\xi$ is the time index modulo $N$~\cite{Alexander2016b}. The decoupling of quantum wires is shown in  Fig.~\ref{fig:wirebsl}(a). These wires are embedded versions of dual-rail wires~\cite{Yokoyama2013, Alexander2014}, as shown in Fig.~\ref{fig:wirebsl}(b).

Input states can be injected into or removed from the BSL using the switching device at ``$\star$" in Fig.~\ref{fig:bslconstruct}(a). When embedded within the resource state, they reside in the ``+" subspace of macronodes on the leftmost edge of the BSL. 
Phase-space displacements can be applied at the ``$\star$" location, though this is equivalent to adapting the later measurements. We show this in Sec.~\ref{sec:adapt}. 

\emph{Measurement constraints}.---Note that measurements at sites `b'~and `c'~are fixed because they are used to decouple the wires~\cite{Alexander2016b}. Therefore, only measurement degrees of freedom at~`x' and~`a' are available to implement single-mode gates. 

\emph{Single-mode Gaussian unitaries}.---Using the homodyne degrees of freedom both~`a' and~`d' (choosing the appropriate setting at `x') implements a single-mode Gaussian unitary gate. Measuring $\hat{p}_{\text{d}}(\theta_{1})$ and $\hat{p}_{\text{a}}(\theta_{2})$ , with outcomes~$m_1$ and~$m_2$, respectively, at a site with an encoded input state implements the Gaussian unitary~\cite{Alexander2016b}
\begin{align}
\label{eq:Vgate}
&
\hat{V}(\theta_{1}, \theta_{2}, m_{1}, m_{2})
\\
&
\coloneqq
	\hat D
	\left[
		\frac {-i e^{i\theta_2} m_1 - i e^{i\theta_1} m_2} {\sin (\theta_1 - \theta_2)}
	\right]
\nonumber 
	\hat{R}(\theta_{+})\hat{S}\left(\ln \tan{\theta_{-}} \right) \hat{R}(\theta_{+}),
\end{align}
where $\theta_{\pm} \coloneqq \frac 1 2 (\theta_1 \pm \theta_2)$, and
$\hat D(\alpha) \coloneqq \exp(\alpha \hat a^\dag - \alpha^* \hat a)$
is a displacement operator.
Up to phase-space displacements, these gates can be used to generate arbitrary single-mode Gaussian unitaries~\cite{Ukai2010, Alexander2014}.

\emph{Single-mode non-Gaussian unitaries}.---Setting detector `x' to measure at `e' and `f' can result in a non-Gaussian unitary operation.  We set detectors `f'~and  `a' to measure in the $\hat{q}$ basis, and we set `e'  to measure $\hat{p}(\theta)$. 
Recall that we can treat the BSL as a collection of disjoint dual-rail wires (see Fig.~\ref{fig:wirebsl}). 

\begin{figure} 
\includegraphics[width=0.8\linewidth]{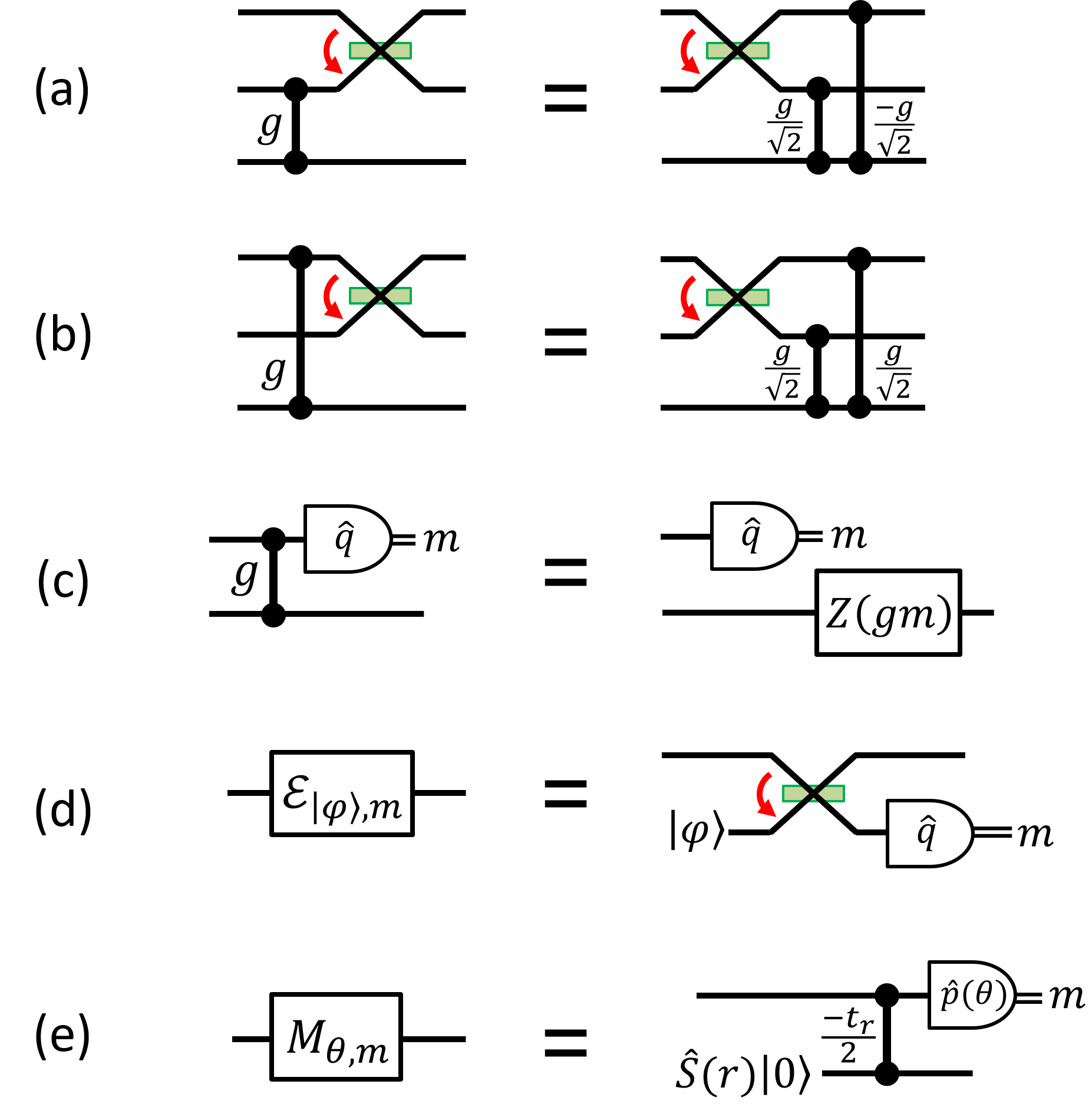}
\caption{ ($\mathbf{a}$)~Commutation of the $\hat{C}_{Z}$ gate through $\op B$ [Eq.\eqref{eq:Cz1}]. ($\mathbf{b}$)~Commutation of the $\hat{C}_{Z}$ gate through $\op B^\dag$ [Eq.\eqref{eq:Cz1}]. ($\mathbf{c}$)  $\hat{q}$-measurement after the $\hat{C}_{Z}$ gate [Eq.\eqref{eq:Cz2}]. ($\mathbf{d}$)~Circuit implementation of the $\mathcal E$ operation [Eq.~\eqref{eq:Eidentity}]. ($\mathbf{e}$)~Measurement-based teleportation identity [Eq.~(\ref{eq:Mdef})].} \label{fig:idents}
\end{figure}

In order to better analyze the effect of this measurement apparatus, in Fig.~\ref{fig:idents} we introduce some useful circuit identities. 
First, consider the following identities involving the $\hat{C}_{Z}$ gate (that follow from the gate definitions): 
\begin{align}
\hat{B}^{(\dagger)}_{jk}\hat{C}_{Z, ij}(g)  &=  \hat{C}_{Z, ij}\left(\frac{g}{\sqrt{2}}\right) \hat{C}^{(\dagger)}_{Z,  ik }\left(\frac{g}{\sqrt{2}}\right)\hat{B}^{(\dagger)}_{jk},\label{eq:Cz1}\\
_{q_{j}}\!\!\bra{s} \hat{C}_{Z,ij}(g) &= \hat{Z}_{i}(s g)\, _{q_{j}}\!\!\bra{s}
,
\label{eq:Cz2}
\end{align}
shown in Fig.~\ref{fig:idents}(a)--(c).  
Next, we define the operation
\begin{align}
	\mathcal{E}_{\ket{\varphi}, m}\coloneqq  
\hat{X} (-m)  \hat{S} \left(\ln \sqrt{2}\right)  \varphi\left(m\sqrt{2} -  \hat{q} \right), ~\label{eq:gatetel} 
\end{align}
where $\varphi(s)= {}_{q\!\!}\braket{s\vert\varphi}$. In Appendix~\ref{sec:ngproof} we show that $\mathcal{E}_{\ket{\varphi}, m}$ can be implemented using the circuit shown in Fig.~\ref{fig:idents}(d), i.e., 
\begin{align}
\label{eq:Eidentity}
	_{q_{k}}\!\!\bra{m} \op B_{jk} \ket \psi_j \otimes \ket \varphi_k
&
=
	 \mathcal{E}_{\ket{\varphi}, m} \ket \psi_j 
	.
\end{align}
 (Note that this is a trace-decreasing map, so for any particular outcome~$m$, the state must be renormalized afterward.)  Finally, Fig.~\ref{fig:idents}(e) shows a \emph{measurement-based teleportation circuit}~\cite{Alexander2014}. In the infinite-squeezing limit,  this 
implements
 \begin{align}
 \label{eq:Mdef}
 \hat{M}_{\theta, m}\coloneqq \hat{X}\left(-2m  \sec\theta \right) \hat{R}\left( -\frac{\pi}{2}\right) \hat{S}\left(\ln \frac{1}{2}\right)  \hat{P}(\tan\theta).
 \end{align}
We provide a proof of Eq.~(\ref{eq:Mdef})  in Appendix \ref{sec:basicmbqc}.

\begin{figure} 
\includegraphics[width=\linewidth]{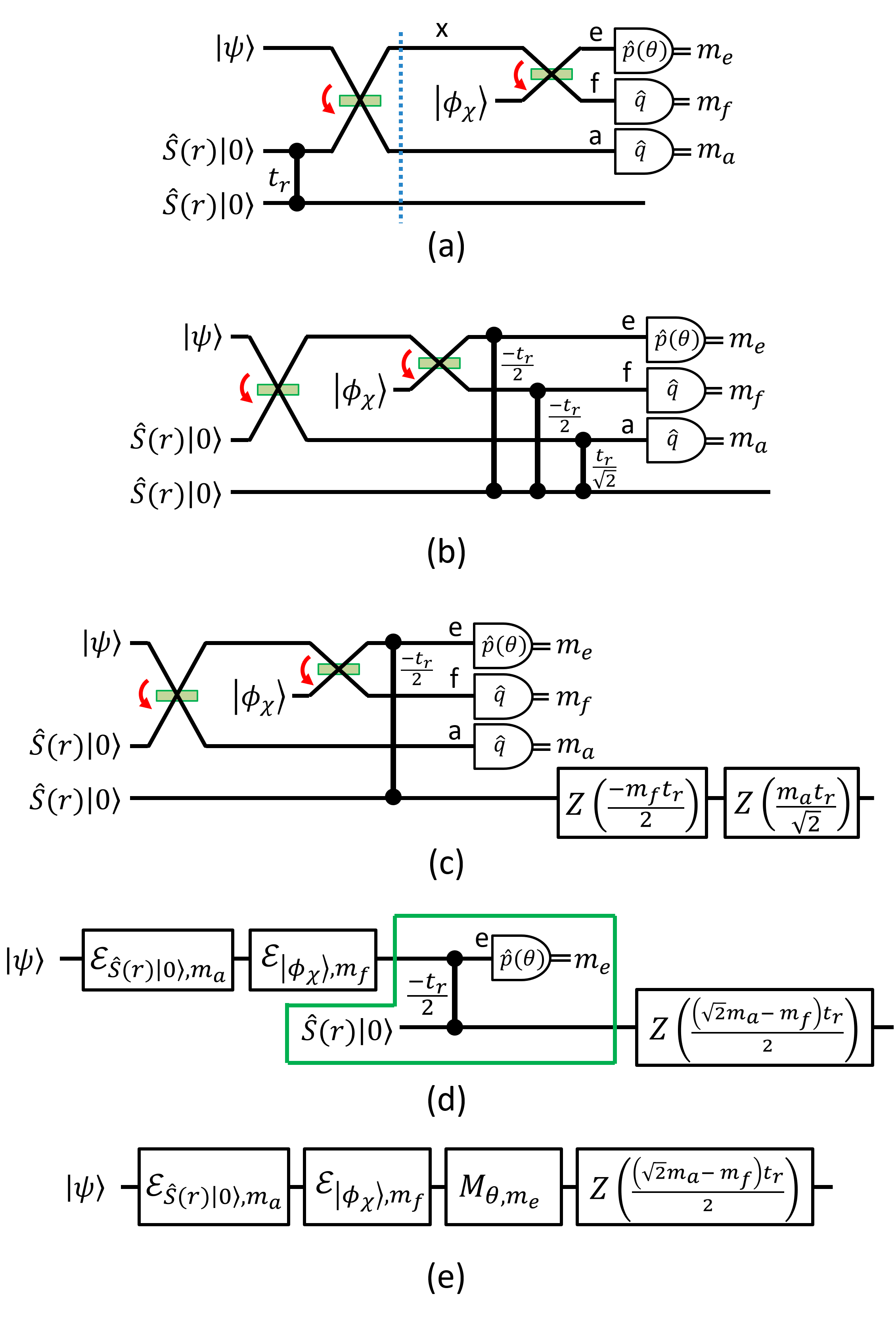}
\caption{$\mathbf{(a)}$~This circuit is equivalent to performing non-Gaussian measurement on the BSL. For compactness, the conditional phase delay shown in Fig.~\ref{fig:bslconstruct}(a) has been incorporated into the homodyne detector at `e'. The input $\ket{\psi}$ should be interpreted as occupying the symmetric subspace of modes `x' and `a' [Eq.~\ref{eq:modelabs}]. Left of the dotted blue line, this circuit is equivalent to the macronode measurement circuit for a dual-rail wire~\cite{Alexander2014}. 
 The parameter $t_{r}=\tanh{2r}$ sets the strength of the  $\hat{C}_{Z}(t_{r})$ gate.  $\mathbf{(b)}$~Starting from panel (a), commute the $\hat{C}_{Z}$ gate rightwards past one beamsplitter [using  Fig.~\ref{fig:idents}(b)] and then past the other [using  Fig.~\ref{fig:idents}(c)]. The end result is three $\hat{C}_{Z}$ gates as shown. $\mathbf{(c)}$~Two of these $\hat{C}_{Z}$ gates precede a measurement in the $\hat{q}$ basis. We can use Fig.~\ref{fig:idents}(d) to replace these with $\hat{Z}$ gates.  $\mathbf{(d)}$~The green box is the measurement-based teleportation circuit [Eq.~\eqref{eq:Mdef}]. $\mathbf{(e)}$~Final equivalent circuit, which implements~$\hat L(\chi, \sigma, \vec m)$ [Eq.~\eqref{eq:equivcircuit}].} \label{fig:nggate}
\end{figure}

Now we analyze the measurement apparatus. Performing our non-Gaussian measurement procedure is equivalent to the quantum circuit shown in Fig.~\ref{fig:nggate}(a). This can be simplified by applying the above identities as described by Figs.~\ref{fig:nggate}(b)-(e). Fig.~\ref{fig:nggate}(e) shows that the measurement implements the following operation:

\begin{align}
\label{eq:equivcircuit}
 \hat{L}(\chi, \sigma, \mathbf{m})  & \coloneqq \nonumber \\
&\hspace{-1.5cm}\hat{Z}\left(\frac{t_{r}\left(\sqrt{2}m_{a}-m_{f}\right)}{2}\right)\hat{M}_{(\tan^{-1}\sigma), m_{e}}  \mathcal{E}_{\ket{\phi_{\chi}}, m_{f}}  \mathcal{E}_{\hat{S}(r)\ket{0}, m_{a}}.
\end{align}
By performing some straightforward algebraic manipulations and taking the infinite squeezing limit, $\hat{L}$ can be simplified further to
 \begin{align}
\hat{L}(\chi, \sigma, \mathbf{m}) & = \nonumber \\
& \hspace{-1cm}\hat{Z}\left(\sqrt{2}m_{a}\right)\hat{X}(\kappa)\hat{R}\left( - \frac{\pi}{2}  \right)
\hat{P}\left(\tau\right) \hat{K}\left(-2\sqrt{2}\chi\right)
,
\label{eq:Kgate}
\end{align}
where 
\begin{align}
\mathbf{m} &\coloneqq (m_{a}, m_{e}, m_{f})
,
\\
\tau &\coloneqq 4\sigma+4 \chi \left(m_{a} +\sqrt{2} m_{f}\right), \label{eq:taudef}
\\
\label{eq:kappadef}
\kappa &\coloneqq -  2 m_{e}  \sqrt{1 + \sigma^2} -2 \sigma \left({ \sqrt{2}m_{a} + m_{f}}\right)
\nonumber \\ & \qquad
- \sqrt{2}\chi \left(m_{a}+\sqrt{2}m_{f} \right)^{2},
\end{align}
and we have neglected the overall phase. We have included a step-by-step proof in Appendix~\ref{sec:cubicphaseproof} for completeness.

The key feature of Eq.~(\ref{eq:Kgate}) is the cubic part, $\hat{K}(\chi)$, which extends our scheme beyond merely Gaussian quantum computation.

 By sequentially applying the measurements described above to a wire [as in Fig.~\ref{fig:wirebsl}(b)], we can generate arbitrary single-mode unitary gates~\cite{Lloyd1999, Gu2009}. Note that with bounded squeezing resources and without quantum error correction, the effective length of the possible quantum computation is bounded from above by a constant~\cite{Ohliger2012}. However, by using encoded qubits~\cite{Gottesman2001}, arbitrarily long qubit-level quantum computation is possible provided that the overall squeezing levels are sufficiently high and the effective qubit noise model is compatible with viable fault-tolerant quantum-error-correction  strategies~\cite{Menicucci2014}. 
 
  Next we show how to implement entangling operations using homodyne measurements. 
\\

\subsection{Entangling operations}

On the BSL, it is possible to implement entangling operations using homodyne detectors~\cite{Alexander2016b}. Neighboring wires are naturally coupled by the BSL graph structure unless deletion measurements are applied, as described in the previous section. 

By measuring a few columns of BSL modes, it is possible to implement Gaussian unitary gates that interact many modes at once. Because the form of such gates can be rather complicated (generally involving many modes and measurement angles), it is convenient to focus on the two-mode case, which is sufficient for demonstrating universality. 

\begin{figure}
\includegraphics[width=\linewidth]{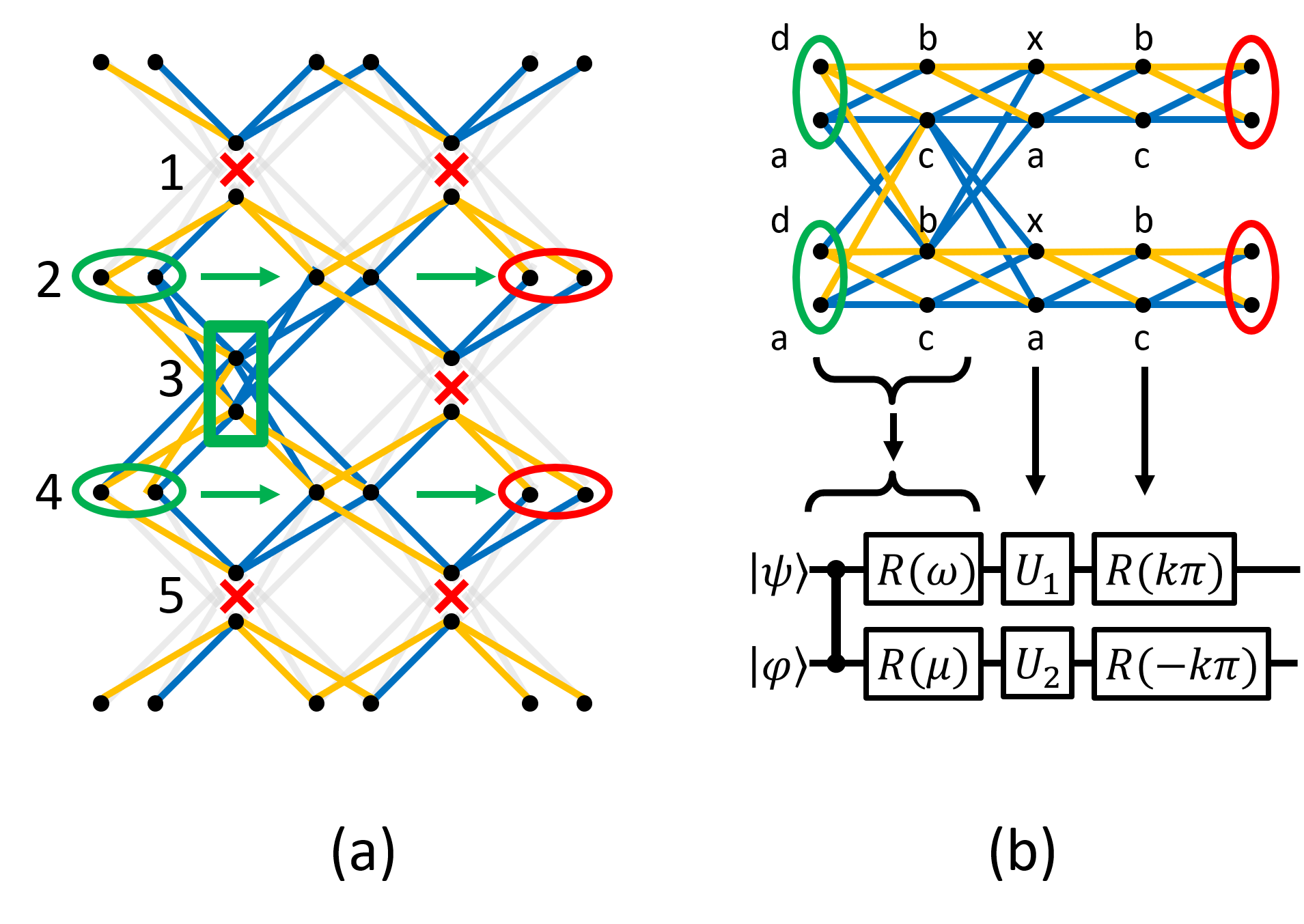}
\caption{$\mathbf{(a)}$ Sublattice containing a region used to implement an entangling gate by measurement-based quantum computation. Detectors `b' and `c' measure $\hat{q}[(-1)^{n} \frac{\pi}{4}]$ at sites marked with a red `X', thereby deleting the grayed-out links. Measurement of macronodes 2, 3, and 4, is described in the main text. The measurements on all other modes can be chosen freely. $\mathbf{(b)}$ After the deletion in panel (a), the middle six rows of modes are equivalent to this subgraph. Measurements on this resource state can implement circuits as shown. Here, the $\hat{C}_{Z}(g)$ gate has weight $g=2\cot{\phi}$. Also, $\omega =  (-1)^{k+1}  \tfrac{3\pi}{4}$ and $\mu =  (-1)^{k}  \tfrac{\pi}{4}$. }
\label{fig:entanglebsl}
\end{figure}

Consider the portion of the BSL shown in Fig.~\ref{fig:entanglebsl}(a). Performing measurements of  $[\hat{p}_{\alpha}(\theta_{2\alpha}), \hat{p}_{\beta}(\theta_{2\beta})]$, $[\hat{p}_{\alpha}(\theta_{3\alpha}), \hat{p}_{\beta}(\theta_{3\beta})]$, and $[\hat{p}_{\alpha}(\theta_{4\alpha}), \hat{p}_{\beta}(\theta_{4\beta})]$ on macronodes 2, 3, and 4, respectively, implements a sequence of beamsplitter and $\hat{V}$ gates [see Eq.~(\ref{eq:Vgate})],
\\

\begin{align}
&\hat{B}_{2+, 4+} \hat{V}_{2+} \left[ (-1)^{k}\frac{\pi}{4}, \theta_{3\alpha}, m_{1\beta}, m_{3\alpha} \right] 
\nonumber \\
& \qquad \times \hat{V}_{4+}\left[\theta_{3\beta},(-1)^{k}\frac{\pi}{4}, m_{3\beta}, m_{5\alpha} \right]
\nonumber \\
& \qquad \times
\hat{B}_{2+, 4+}\hat{V}_{2+}(\theta_{2\alpha}, \theta_{2\beta}, m_{2\alpha}, m_{2\beta} )
\nonumber \\
& \qquad \times
\hat{V}_{4+}(\theta_{4\alpha}, \theta_{4\beta}, m_{4\alpha}, m_{4\beta} )  \label{eq:gentmgate}
\end{align}
on inputs initially encoded within the symmetric subspace of macronodes 2 and 4 (denoted by the subscripts 2+ and 4+, respectively)~\cite{Alexander2016b}.  Above, $k$ denotes the time index of macronode 2.  Next, we consider two examples of gates included in this class.

If $\theta_{3\alpha} = \theta_{3\beta} = (-1)^{k+1}\frac{\pi}{4} $ in Eq.~(\ref{eq:gentmgate}) then up to displacements and overall phase,
\begin{align}
\left(  \hat{V}\otimes\hat{V}  \right)  \hat{B}= \hat{B}^{\dagger} \left(  \hat{V}\otimes\hat{V} \right) 
\end{align}
(we omit subscripts and dependence on measurement variables). Thus, Eq.~(\ref{eq:gentmgate}) is reduced to two copies of the single-mode gate case described in the previous section.

By brute-force search we found that choosing 
\begin{align}
\begin{pmatrix} \theta_{2\alpha} \\ \theta_{2\beta} \\ \theta_{3\alpha} \\ \theta_{3\beta} \\ \theta_{4\alpha} \\ \theta_{4\beta}\end{pmatrix}
 =\begin{pmatrix} (-1)^{k+1} \frac{\pi}{8} \\ (-1)^{k} \frac{3\pi}{8} \\ (-1)^{k} \frac{\pi}{4} + \phi \\ (-1)^{k} \frac{\pi}{4} - \phi \\(-1)^{k+1} \frac{\pi}{8} \\ (-1)^{k} \frac{3\pi}{8}\end{pmatrix}
\end{align}
simplifies the gate in Eq.~(\ref{eq:gentmgate}) to a gate of the form
\begin{align}
& \left[\hat{X}( \lambda_{1} ) \hat{Z}( \lambda_{2} ) \otimes \hat{X}( \lambda_{3}  ) \hat{Z}( \lambda_{4} )\right] \nonumber \\ 
& \times
\left[\hat{R}\left(  (-1)^{k+1}  \frac{3\pi}{4} \right) \otimes \hat{R}\left(  (-1)^{k}  \frac{\pi}{4} \right)\right] \hat{C}_{Z}(2\cot{\phi}) \label{eq:Czgate}
\end{align}
for some $\lambda_{1}$, $\lambda_{2}$, $\lambda_{3}$, and $\lambda_{4}$~\cite{Alexander2016b}.

Note that this implements a $\hat{C}_{Z}$ gate (of variable weight) up to a pair of local phase delays. These phase delays can be undone using measurements further along the BSL, such as with gates $\hat{U}_{1}$ and $\hat{U}_{2}$ in Fig.~\ref{fig:entanglebsl}(b). 

\subsection{Adaptivity of the measurement outcomes}\label{sec:adapt}

Above we described how to implement a universal set of gates [see Eqs.~(\ref{eq:Vgate}), (\ref{eq:Kgate}), and (\ref{eq:Czgate})]. In each case, however,  the gate has only been implemented up to a known phase-space  displacement that  depends  on the random measurement outcomes. To make computation deterministic, one can apply appropriate displacements at the ``$\star$" site in Fig.~\ref{fig:bslconstruct}(a) or use an adaptive measurement  protocol---i.e., later measurement bases are chosen in a way that depends on the values of prior measurement outcomes~\cite{Gu2009}. 

Measurement protocols consisting only of homodyne measurements are an exception. They can be implemented deterministically without such adaptivity, though these protocols are not universal and only implement Gaussian unitary gates~\cite{Menicucci2006, Gu2009}. To see this, note that the phase-space displacements form a normal subgroup of the Gaussian unitaries. Therefore, the randomness of the measurement outcomes only causes Gaussian computations to differ up to a final phase-space displacement.  
This in turn can then be dealt with by classically processing the final homodyne measurement data.  

We now consider the case of implementing $\hat{L}(\chi, \sigma, \mathbf{m})$ from Eq.~\eqref{eq:Kgate}. Suppose that $\hat{X}(s)$ and $\hat{Z}(t)$ are unwanted  phase-space displacements resulting from the previous step of MBQC. Now we consider commuting these so that they act after $\hat{L}(\chi, \sigma, \mathbf{m})$:
\begin{align}
&\hat{L}(\chi, \sigma, \mathbf{m}) \hat{Z}(t) = \hat{X}(t) \hat{L}(\chi, \sigma, \mathbf{m}),\\
&\hat{L}(\chi, \sigma, \mathbf{m}) \hat{X}(s) \nonumber \\
& \quad = \hat{Z}(-s) \hat{X}(  \tau s  - 2 \sqrt 2 s^2 \chi )  
\hat{L}\left(\chi, \sigma, \mathbf{m}\right)\hat{P}( -4\sqrt{2} s \chi )
,\label{eq:LXcom} 
\end{align}
where we neglect overall phases. Note that in the case of $\hat{Z}(t)$, commutation still results in a displacement operator. The $\hat{X}(s)$ case is more complicated, generating additional position and momentum displacements, as well as a shear. By adapting the homodyne measurement  angle~$\theta = \tan^{-1} \sigma$  in Eq.~(\ref{eq:LXcom}), we can cancel out the effect of the shear $\hat{P}( -4 \sqrt{2} s \chi)$ at the price of an additional contribution to the final phase-space displacement. Modifying the shear parameter $\sigma\rightarrow \sigma^{\prime} = \sigma +  \sqrt{2} s \chi $ in Eq.~(\ref{eq:LXcom}), equivalent to $\theta \to \theta' = \tan^{-1}(\tan \theta +\sqrt{2} s \chi )$, and applying the appropriate commutation relations, we get
\begin{align}
\hat{L}(\chi, \sigma^{\prime}, \mathbf{m}) \hat{X}(s)  &=   \hat{Z}(-s)  \hat{X}\left(\zeta \right) \hat{L}\left(\chi, \sigma, \mathbf{m}\right),
\label{eq:LXcom2}
\end{align}
where 
\begin{align}
&\zeta \coloneqq  4 s \sigma + 2\sqrt{2} s \chi (m_{f}+s) 
- 2 m_{e}(\sqrt{1+{\sigma^\prime}^{2}}- \sqrt{1+\sigma^{2}}). 
\end{align}

In other words, by making detector `e' (or equivalently, the parameter $\sigma$) adaptive, we can compensate control for random phase-space displacements and deterministically implement non-Gaussian unitary gates, up to a final known phase space displacement. This is the only type of adaptive measurement required by our protocol since unwanted displacements at the end of the computation can be dealt with by post-processing the measurement data~\cite{Gu2009}.

Our approach is very similar to the adaptive homodyne techniques used in Ref.~\cite{Miyata2016} for  cubic-phase-gate teleportation. Implementing adaptive squeezing operations (e.g., at the location ``$\star$") is experimentally infeasible, so it is significant that our scheme only requires adaptive linear optics.

 \section{Conclusion}\label{sec:conc}
Here we have proposed a method using temporal modes for generating the bilayer-square-lattice cluster state---a universal resource for measurement-based quantum computation~\cite{Alexander2016b}. Our scheme only requires four sources of squeezed vacuum modes (such as an optical parametric oscillator) and a few beamsplitters. The simplicity of this approach makes it a natural two-dimensional generalization of one-dimensional resource states generated in Refs.~\cite{Yokoyama2013, Yoshikawa2016}.

We showed by using properties of the bilayer square lattice's graph that it is equivalent under local phase delays to a Gaussian pure state that has essentially the same graph and possesses approximate local nullifiers composed purely of either position or momentum operators. The verification of genuine multipartite entanglement on cluster states that are equivalent to pairwise-squeezed states is experimentally straightforward:\ all modes are simply measured in the position and momentum basis and appropriate linear combinations are taken, whose variance is compared to an appropriate entanglement witness~\cite{vanLoock2003}. Furthermore, these features are shared by an entire family of cluster states that approximate trace-zero, self-inverse, ideal (i.e., infinitely-squeezed) cluster states.

Our proposal extends previous work by explicitly incorporating non-Gaussian elements into the measurement devices~\cite{Ukai2010, Ukai2011, Su2013, Miyata2014, Ferrini2013a, Alexander2014, Yokoyama2015, Ferrini2016, Alexander2016}. Such elements enable universal quantum computation. Our approach conveniently minimizes the requirements of measurement adaptivity, potentially reducing the noise due to finite squeezing, although a proper analysis of this is left to future work.  One additional advantage of the chosen gate set is that it is readily compatible with the universal gate set for the \emph{Gottesman-Kitaev-Preskill} qubit~\cite{Gottesman2001}, which is a key ingredient in the proof of fault-tolerant quantum computation using continuous-variable cluster states~\cite{Menicucci2014}.

\acknowledgments

This work was partially supported by National Science Foundation Grant {No.\ PHY-1630114}. S. Y.\ acknowledges support from JSPS Overseas Research Fellowships of Japan and the Australian Research Council Centre of Excellence for Quantum Computation and Communication Technology (Grant number CE110001027). A.F.\ acknowledges support from Core Research for Evolutional Science and Technology (Grant No.\ JPMJCR15N5) of Japan Science and Technology Agency, Grants-in-Aid for Scientific Research of Japan Society for the Promotion of Science, and Advanced Photon Science Alliance. N.C.M.\ is supported by the Australian Research Council Centre of Excellence for Quantum Computation and Communication Technology (Project No.\ CE170100012).

\bibliography{TMBSL.bbl}

\newpage

\appendix

\section{Proof of Theorem~\ref{thm:1}}\label{sec:thmproof}

The theorem contains two claims. We prove them in order. Our proof of the first claim proceeds in three steps.
\begin{enumerate}[label=(\roman*),itemsep=4pt]
\item We directly construct a Gaussian pure state~$\ket{\Psi_r}$ with graph~$\mat Z_{\Psi,r}$ from the ideal graph~$\mat V$ and  prove that it meets the definition of an approximate CV cluster state whose ideal graph is~$\mat V$.
\item We construct $\ket{\Phi_r}$ by applying phase delays to $\ket{\Psi_r}$ and prove that we can define a set of approximate nullifiers for this state (that are exact in the infinite-squeezing limit) that contain position or momentum only (and no combinations of the two).
\item We construct new linear combinations of these nullifiers, without mixing position and momentum, and prove that this new set of nullifiers has its support limited to the neighborhood (as defined by~$\mat V$) of some particular node.
\end{enumerate}
We define~$\ket{\Psi_r}$ to be the Gaussian pure state whose graph is~\cite{Menicucci2011}
\begin{align}
\label{eq:ZPsi}
	\mat Z_{\Psi,r} \coloneqq i (\sech 2r) \mat \id_{2n} + (\tanh 2r) \mat V.
\end{align}
It follows that the set~$\{ \mat Z_{\Psi,r} | r > 0\}$ is a family of Gaussian pure states indexed by a real parameter~$r$ (the overall squeezing parameter) such that
\begin{align}
	\lim_{r\to\infty} \mat Z_{\Psi,r} = \mat V.
\end{align}
Any member of this set meets the definition~\cite{Menicucci2011} of an approximate CV cluster state with an ideal graph~$\mat V$. Therefore, $\ket{\Psi_r}$ is an approximate CV cluster state with an ideal graph~$\mat V$. This proves step~(i).

Let
\begin{align}
	\op R_\theta \coloneqq \op R(\theta)
\end{align}
for brevity. We now define
\begin{align}
	\ket{\Phi_r}
&
\coloneqq
	\op R_{\pi/4}^{\otimes 2n} \ket {\Psi_r}.
\end{align}
Using the fact that
\begin{align}
	\bigl(\op R_{\pi/4}^{\otimes 2n}\bigr)^\dagger \opvec x \bigl(\op R_{\pi/4}^{\otimes 2n}\bigr)
&
=
	\frac {1} {\sqrt 2}
	\begin{pmatrix}
		\mat I_{2n}	& -\mat I_{2n}
	\\
		\mat I_{2n}	& \mat I_{2n}
	\end{pmatrix}
	\opvec x
	,
\end{align}
then the rule for updating the graph for a Gaussian pure state~\cite{Menicucci2011} gives
\begin{align}
\label{eq:ZPhi}
	\mat Z_{\Phi,r}
&
=
	(\mat I_{2n} + \mat Z_{\Psi,r})(\mat I_{2n} - \mat Z_{\Psi,r})^{-1}.
\end{align}
To simplify this, we note that, from the assumptions of the theorem, $\mat V = \mat V^\tp = \mat V^{-1} \in \reals^{2n \times 2n}$ and $\exists \mat L \in \group O(2n)$ such that
\begin{align}
\label{eq:Vdiag}
	\mat V = \mat L (\mat I_n \oplus -\mat I_n) \mat L^\tp,
\end{align}
where $\oplus$ represents the matrix direct sum (i.e., it creates a block-diagonal matrix). This particular form is guaranteed because $\tr \mat V = 0$ and all of $\mat V$'s eigenvalues must be~$\pm 1$. For brevity later, we also define
\begin{align}
\label{eq:zpmdef}
	z_\pm
&
\coloneqq
	 i \sech 2r \pm \tanh2r,
\end{align}
for which we have the following identity:
\begin{align}
\label{eq:zpm}
	\frac {1+z_\pm} {1-z_\pm} = i e^{\pm 2r}.
\end{align}
Using Eqs.~\eqref{eq:Vdiag} and~\eqref{eq:zpmdef}, we can rewrite Eq.~\eqref{eq:ZPsi} as
\begin{align}
\label{eq:ZPsisimple}
	\mat Z_{\Psi,r}
&
=
	\mat L( z_+ \mat I_n \oplus z_- \mat I_n ) \mat L^\tp.
\end{align}
Plugging Eq.~\eqref{eq:ZPsisimple} into
Eq.~\eqref{eq:ZPhi} and using Eq.~\eqref{eq:zpm} gives
\begin{align}
\label{eq:ZPhisimple}
	\mat Z_{\Phi,r}
&
=
	i\mat L (e^{2r}\mat I_n \oplus e^{-2r} \mat I_n) \mat L^\tp.
\end{align}
Since~$\mat Z_{\Phi,r}$ is purely imaginary, we already know~\cite{Menicucci2011} that it contains only $q$-$q$ and $p$-$p$ correlations (i.e., no $q$-$p$ correlations). But we still need to calculate the nullifiers.

The exact nullifiers for~$\mat Z_{\Phi,r}$ can be obtained from the usual relation~\cite{Menicucci2011}
\begin{align}
\label{eq:nullPhir}
	\vec 0 &= (\opvec p - \mat Z_{\Phi,r} \opvec q) \ket {\Phi_r},
\intertext{which, after left multiplication by~$-\mat Z_{\Phi,r}^{-1}$, also implies}
\label{eq:invnullPhir}
	\vec 0 &= (-\mat Z_{\Phi,r}^{-1} \opvec p + \opvec q) \ket {\Phi_r}.
\end{align}
For brevity later on, let us denote the top and bottom halves of~$\mat L^\tp$ by
\begin{align}
	\mat L_+^\tp
&
\coloneqq
	\begin{pmatrix}
		\mat I_n	& \mat 0
	\end{pmatrix}
	\mat L^\tp,
\\
	\mat L_-^\tp
&
\coloneqq
	\begin{pmatrix}
		\mat 0	& \mat I_n
	\end{pmatrix}
	\mat L^\tp.
\end{align}
Multiplying Eqs.~\eqref{eq:nullPhir} and~\eqref{eq:invnullPhir} on the left by $\mat L_-^\tp$ and~$\mat L_+^\tp$, respectively, gives
\begin{align}
	\vec 0
&
=
	\mat L_-^{\text{T}}
	(\opvec p - i e^{-2r} \opvec q) \ket {\Phi_r},
\\
	\vec 0
&
=
	\mat L_+^\tp
	(i e^{-2r} \opvec p + \opvec q) \ket {\Phi_r}.
\end{align}
In the limit~$r \to \infty$, we get
\begin{align}
\label{eq:qpnull}
	\begin{pmatrix}
		\mat L_-^\tp \opvec p
	\\
		\mat L_+^\tp \opvec q
	\end{pmatrix}
	\ket {\Phi_r}
	\to
	\vec 0
	.
\end{align}
This proves step~(ii).

The final step is to find linear combinations of these nullifiers that (a)~do not mix $\op q$ and $\op p$ and (b)~are local with respect to~$\mat V$. The neighborhood of node~$j$ with respect to~$\mat V$ is given by the nonzero entries of the $j$th row (or, equivalently, the $j$th column) of~$\mat V$.

Examining the structure of~$\mat V$ shown in Eq.~\eqref{eq:Vdiag}, we see that
\begin{align}
	\mat V
&
=
	\mat L_+ \mat L_+^\tp
	-
	\mat L_- \mat L_-^\tp
.
\end{align}
Therefore, since $\mat L_\pm^\tp \mat L_\mp = \mat 0$, we have $\mat I_{2n} = \mat V^2 = \mat L_+ \mat L_+^\tp + \mat L_- \mat L_-^\tp$, and thus,
\begin{align}
	\mat L_\pm \mat L_\pm^\tp
&
=
	\frac 1 2 (\mat I_{2n} \pm \mat V).
\end{align}
Therefore, we can multiply Eq.~\eqref{eq:qpnull} from the left by $2 (\mat L_- \oplus \mat L_+)$ to obtain
\begin{align}
	\begin{pmatrix}
		(\mat I_{2n} - \mat V) \opvec p
	\\
		(\mat I_{2n} + \mat V) \opvec q
	\end{pmatrix}
	\ket {\Phi_r}
	\to
	\vec 0
	.	
\end{align}
This proves step~(iii) and therefore proves the first claim.

To prove the second claim, we simply evaluate Eq.~\eqref{eq:ZPhisimple}:
\begin{align}
\label{eq:}
\mat Z_{\Phi,r}
&
=
	i\mat L (e^{2r}\mat I_n \oplus e^{-2r} \mat I_n) \mat L^\tp
\nonumber \\
&
=
	i \exp(2r \mat V)
\nonumber \\
&
=
	i (\cosh 2r) \mat I_{2n} + i (\sinh 2r) \mat V,
\end{align}
where we use the fact that~$\mat V^2 = \mat I_{2n}$ to obtain the last line.
\hfill \qed

\begin{widetext}
\section{Gate gadget action}\label{sec:ngproof}
Here we directly calculate the effect of the circuit shown in  Fig.~\ref{fig:idents}(d). 
The operation in Fig.~\ref{fig:idents}(d) is given by
\begin{align}
\mathcal{E}_{\ket{\varphi}, m} & \coloneqq   \,_{q_{2}}\!\!\bra{m}  \hat{B}_{1, 2} \int \mathrm{d}s\, \varphi(s) \ket{s}_{q_{2}}. \label{eq:A1} \\ 
&=\int \mathrm{d}s\, \varphi(s)  \,_{q_{2}} \!\!\bra{m} \hat{B}_{1, 2} \int \mathrm{d}t \ket{t}_{p_{1} \, \,p_{1}}\!\!\braket{t\vert s}_{q_{2}}
\end{align}
Using the fact that $\hat{B}_{1, 2} \ket{t}_{p_{1}}\ket{s}_{q_{2}}$ is equivalent to an infinitely squeezed, displaced two-mode squeezed state (equivalent to an EPR state~\cite{Einstein1935}), i.e.,
\begin{align}
\hat{B}_{1, 2} \ket{t}_{p_{1}}\ket{s}_{q_{2}}
&=
\hat{X}_{1}\left( \frac{-s}{\sqrt{2}}\right) \hat{X}_{2}\left(\frac{s}{\sqrt{2}} \right) \hat{Z}_{1}\left(\frac{t}{\sqrt{2}}\right) \hat{Z}_{2}\left(\frac{t}{\sqrt{2}}\right) \frac {1} {\sqrt{\pi}}  \int^{\infty}_{-\infty} \mathrm{d} r\,  \ket{r}_{q_1}\ket{r}_{q_2}
\\
&=
\hat{X}_{1}\left( \frac{-s}{\sqrt{2}}\right) \hat{X}_{2}\left(\frac{s}{\sqrt{2}} \right)  \frac {1} {\sqrt{\pi}} \int^{\infty}_{-\infty} \mathrm{d} r\,  e^{i \sqrt 2 tr} \ket{r}_{q_1}\ket{r}_{q_2}
\\
&=
 \frac {1} {\sqrt{\pi}}  \int^{\infty}_{-\infty} \mathrm{d} r\,  e^{i \sqrt 2 tr}
\left\lvert r - \frac {s} {\sqrt 2} \right\rangle_{q_1} \left\lvert r + \frac {s} {\sqrt 2} \right\rangle_{q_2}
,
\end{align}
we get that
\begin{align}
\mathcal{E}_{\ket{\varphi}, m}
&=
 \frac {1} {\sqrt{\pi}}  \int \mathrm{d}s\,  \varphi(s) \bra{m}_{q_{2}}
\int \mathrm{d}t  \int \mathrm{d} r\,  e^{i \sqrt 2 tr}
\left\lvert r - \frac {s} {\sqrt 2} \right\rangle_{q_1} \left\lvert r + \frac {s} {\sqrt 2} \right\rangle_{q_2}\,_{p_{1}}\!\!\bra{t}\\
&=
 \frac {1} {\sqrt{\pi}}  \int \mathrm{d}s\,  \varphi(s) 
\int \mathrm{d}t  \int \mathrm{d} r\,  e^{i \sqrt 2 tr}
\delta\left( m - r -\frac{s}{\sqrt{2}}\right)
\left\lvert r - \frac {s} {\sqrt 2} \right\rangle_{q_1}  \,_{p_{1}}\!\!\bra{t}\\
&=
  \frac {1} {\sqrt{\pi}}  \int \mathrm{d}s\,  \varphi(s) 
\int \mathrm{d}t \,  
e^{it(\sqrt{2} m - s) }\left\lvert m - \sqrt{2} s \right\rangle_{q_1}  \,_{p_{1}}\!\!\bra{t}
\end{align}
Then, using that 
\begin{align}
\ket{s}_{q}= \frac {1} {\sqrt{2\pi}} \int \mathrm{d}t\, e^{-i t s}\ket{t}_{p},
\end{align}
this simplifies further:
\begin{align}
\mathcal{E}_{\ket{\varphi}, m}
&=
 \sqrt 2  \int \mathrm{d}s \left\vert m - \sqrt{2} s \right\rangle_{q_{1}\,\, q_{1}}\!\!\left\langle \sqrt 2 m - s \right\rvert \varphi(s) 
\\
&=
 \sqrt 2  \hat{X}_{1}(-m) \int \mathrm{d}s \left\vert 2 m - \sqrt{2} s \right\rangle_{q_{1}\,\, q_{1}}\!\!\left\langle \sqrt 2 m - s \right\rvert  \varphi\left( \sqrt{2}  m- \hat{q}_1 \right) 
\\
&= 
  \hat{X}_{1}(-m) \hat{S}_{1}\left(\ln \sqrt{2}\right) \int \mathrm{d}s \left\vert \sqrt 2 m - s \right\rangle_{q_{1}\,\, q_{1}}\!\!\left\langle \sqrt 2  m- s \right\vert  \varphi\left( \sqrt{2} m - \hat{q}_1\right) 
\\
&=
  \hat{X}_{1}(-m) \hat{S}_{1}\left(\ln \sqrt{2}\right)  \varphi\left( \sqrt{2} m - \hat{q}_1\right), 
\end{align}
in agreement with Eq.~(\ref{eq:gatetel}).

\section{Measurement-based circuit}\label{sec:basicmbqc}
Here we review the effective operation implemented by the circuit in Fig.~\ref{fig:idents}(e) for an input state $\ket{\psi_{1}}$. After the $\hat{C}_{Z}$ gate, we consider measuring the top mode in the  $\hat{P}^{\dagger}(\tan{\theta})\hat{p}\hat{P}(\tan{\theta})=\hat{p} + (\tan{\theta}) \hat{q}$ basis, which is equivalent to measuring $\hat{p}(\theta)$, obtaining outcome~$m$, which is multiplied by $\sec{\theta}$ to obtain the effective outcome~$m^{\prime}= m\sec{\theta}$ of the desired measurement~\cite{Menicucci2006}.  The output state $\ket{\psi_{2}}$ is given by 
\begin{align}
\ket{\psi_{2}}_{k}&\propto \, _{p_{j}}\!\!\bra{m^{\prime}}\hat{P}_{j}(\tan{\theta}) \hat{C}_{Z}(g) \ket{\psi_1}_{j} \otimes \left(\exp\left[-\frac{\hat{q}_{k}^{2}}{2 e^{2r}}\right]\ket{0}_{p_{k}} \right) 
\end{align}
Taking the infinite squeezing limit ($r\rightarrow\infty$),
\begin{align}
\lim_{r\rightarrow\infty}\ket{\psi_{2}}_{k} &=\text{Pr}(m^{\prime})^{-\frac{1}{2}} \, _{p_{j}}\!\!\bra{m^{\prime}} \hat{C}_{Z}(g)  \ket{\psi_1^{\prime}}_{j}  \otimes\ket{0}_{p_{k}}.
\end{align}
where $\ket{\psi_{1}^{\prime}} \coloneqq \hat{P}(\tan{\theta}) \ket{\psi_{1}}$ and where $\Pr(x)$ is the probability of outcome~$x$ in the infinite squeezing limit. 
 Next, we use squeezers and $\pi$-phase delays to convert the $\hat{C}_{Z}$ gate weight from $g\mapsto 1$ 
\begin{align}
\lim_{r\rightarrow\infty}\ket{\psi_{2}}_{k}&=\text{Pr}(m^{\prime})^{-\frac{1}{2}} \, _{p_{j}}\!\!\bra{m^{\prime}}  \hat{S}_{k}^{\dagger}(\ln \abs{g})\hat{R}^{\dagger}_{k}(\Im \ln g)\hat{C}_{Z}(1) \hat{R}_{k}(\Im \ln g) \hat{S}_{k}(\ln \abs g )   \ket{\psi_1^{\prime}}_{j}  \otimes\ket{0}_{p_{k}} \\
&=\text{Pr}(m^{\prime})^{-\frac{1}{2}}  \hat{S}_{k}\left( \ln \frac{1}{\abs g} \right) \hat{R}^{\dagger}_{k}(\Im \ln g) \, _{p_{j}}\!\!\bra{m^{\prime}}  \hat{C}_{Z}(1)   \ket{\psi_1^{\prime}}_{j} \otimes\ket{0}_{p_{k}}. \label{eq:workingpsi2}
\end{align}
The final part of this expression is the standard weight-1 canonical continuous-variable cluster state teleportation circuit~\cite{Menicucci2006, Gu2009}. The output is well known:
\begin{align}
\text{Pr}(m^{\prime})^{-\frac{1}{2}}\, _{p_{j}}\!\!\bra{m^{\prime}}  \hat{C}_{Z}(1) \ket{\psi_1^{\prime}}_{j} \otimes\ket{0}_{p_{k}}=\hat{X}_{k} (m^{\prime})\hat{R}_{k}\left(\frac{\pi}{2}\right) \ket{\psi_1^{\prime}}_{k}. \label{eq:cancvcs}
\end{align}
Combining Eq.~\eqref{eq:cancvcs} with Eq.~\eqref{eq:workingpsi2}, plugging in $g= -t_{r}/2$, and taking the infinite squeezing limit results in 
\begin{align}
\lim_{r\rightarrow\infty}\ket{\psi_{2}}
&= \hat{S}(\ln 2) \hat{R}^{\dagger}(\pi) \hat{X}(m^{\prime})\hat{R}\left(\frac{\pi}{2}\right) \ket{\psi_1^{\prime}}\\
&= \hat{S}(\ln 2) \hat{X}(-m^{\prime})\hat{R}\left(-\frac{\pi}{2}\right) \ket{\psi_1^{\prime}}\\
&= \hat{X}(-2m^{\prime})\hat{R}\left(-\frac{\pi}{2}\right) \hat{S}\left(\ln \frac 1 2\right) \ket{\psi_1^{\prime}}\\
&= \hat{X}\left(-2m  \sec\theta \right) \hat{R}\left( -\frac{\pi}{2}\right) \hat{S}\left(\ln \frac{1}{2}\right)  \hat{P}(\tan\theta)\ket{\psi_{1}}\\
&=\hat{M}_{\theta, m}\ket{\psi_{1}}
\end{align}
in agreement with Eq.~\eqref{eq:Mdef}.

\section{Step-by-step simplification of $\hat{L}$}\label{sec:cubicphaseproof}

Here we begin with the sequence of operations implemented by  Fig.~\ref{fig:nggate}(e)  and show how it can be simplified to Eq.~(\ref{eq:Kgate}).  For brevity in what follows, we define the following combinations of measurement outcomes:
\begin{align}
	m_\pm \coloneqq \frac {\sqrt{2}m_{a}\pm m_{f}} {2}.
\end{align}
Using this compact notation, we rewrite Eq.~\eqref{eq:equivcircuit} as
\begin{align}
\hat{L}(\chi, \sigma, \mathbf{m}) & = 
\hat{Z}( t_{r} m_- ){\hat{M}_{(\tan^{-1}\sigma), m_{e}} }  \mathcal{E}_{\ket{\phi_{\chi}}, m_{f}}  \mathcal{E}_{\hat{S}(r)\ket{0}, m_{a}}.
\end{align}
For convenience, we repeat the definitions
\begin{align}
\hat{M}_{\theta, m}
&\coloneqq
\hat{X}(-2m \sec\theta) \hat{R}\left(  - \frac{\pi}{2}  \right) \hat{S}\left(\ln \frac{1}{2}\right)  \hat{P}(\tan\theta)
,
 \\
	\mathcal{E}_{\ket{\varphi}, m}
&\coloneqq 
	\hat{X}(-m) \hat{S}\left(\ln \sqrt{2}\right)\varphi\left(\sqrt{2} m-\hat{q}\right) 
,
\label{eq:apchanE}
\end{align}
from Eqs.~\eqref{eq:Mdef} and~\eqref{eq:gatetel}, respectively.  For $\ket{\varphi}= \hat{S}(r)\ket{0}$, the $\varphi$ term in Eq.~\eqref{eq:apchanE} is a normalized Gaussian envelope with variance $e^{2r} / 2$. In the infinite squeezing limit, this term acts trivially and can be ignored.  

Applying the relevant definitions and taking the infinite-squeezing limit
gives
\begin{align}
\hat{L}(\chi, \sigma, \mathbf{m})& \propto  \hat{Z} (m_-)  \hat{X}\left(-2m_{e}  \sqrt{1 + \sigma^2} \right) \hat{R}\left(  - \frac{\pi}{2}  \right) \hat{S}\left(\ln \frac{1}{2}\right)  \hat{P}(\sigma)  \hat{X}(-m_{f}) \hat{S}(\ln \sqrt{2})\phi_{\chi}\left(\sqrt{2} m_{f}-\hat{q}\right) \hat{X}(-m_{a})\hat{S}(\ln \sqrt{2}),
\end{align}
where the $\propto$ sign indicates that we have omitted the overall phase. 
Now, we commute the $\hat{S}(\ln \sqrt{2})$ gates leftwards so they cancel with $\hat{S}\left(\ln \frac{1}{2}\right)$, giving
\begin{align}
\hat{L}(\chi, \sigma, \mathbf{m})& \propto \hat{Z} (m_-)  \hat{X}\left(-2m_{e} \sqrt{1 + \sigma^2} \right) \hat{R}\left( - \frac{\pi}{2}  \right)  \hat{P}(4\sigma)  \hat{X}\left(\frac{-m_{f}}{2}\right) \phi_{\chi}\left(\sqrt{2}\left(m_{f}-\hat{q}\right)\right) \hat{X}\left(\frac{-m_{a}}{\sqrt{2}}\right).
\end{align}
Next, we commute displacements to the left, step by step:
\begin{align}
\hat{L}(\chi, \sigma, \mathbf{m})& \propto  \hat{Z}\left(m_{-}\right)\hat{X}\left(-2m_{e}  \sqrt{1 + \sigma^2} \right) \hat{R}\left( - \frac{\pi}{2} \right)  \hat{P}(4\sigma)  \hat{X}\left(-m_{+}\right) \phi_{\chi}\left(\sqrt{2}\left(m_{f}-\hat{q}\right) + m_{a} \right)  \\
& \propto  \hat{Z}\left(m_{-}\right)\hat{X}\left(-2m_{e}  \sqrt{1 + \sigma^2} \right) \hat{R}\left(  - \frac{\pi}{2}  \right)\hat{X}\left(-m_{+}\right)  \hat{Z}\left(-4\sigma m_{+}\right) \hat{P}(4\sigma)  
 \phi_{\chi}\left(\sqrt{2}\left(m_{f}-\hat{q}\right) + m_{a} \right)\\
& \propto  \hat{Z}\left(\sqrt{2}m_{a}\right)\hat{X}\left[-2m_{e}  \sqrt{1 + \sigma^2} - 2 \sigma \left(  \sqrt{2} m_{a} + m_{f} \right)\right] \hat{R}\left( -\frac{\pi}{2} \right)  \hat{P}(4\sigma)  \phi_{\chi}\left(\sqrt{2}\left(m_{f} -\hat{q}\right) + m_{a}\right) \label{eq:Lsimpphi}
\end{align}

 Using $\phi_{\chi}(s) = e^{i \chi s^{3}/3} $, the rightmost term in Eq.~(\ref{eq:Lsimpphi}) can be expanded  as
 \begin{align}
 \phi_{\chi}\left(\sqrt{2}\left(m_{f} - \hat{q} \right) + m_{a}\right) &= e^{i \frac{\chi}{3} \left(\sqrt{2}\left(m_{f} -\hat{q}\right) + m_{a}\right)^{3}} \\
 &= \text{(phase)}\,  e^{-i \sqrt{2}\chi(m_{a}+ \sqrt{2} m_{f})^{2} \hat{q}} e^{i (2\chi m_{a} + 2 \sqrt{2} \chi m_{f})\hat{q}^{2}} e^{-i \frac{2\sqrt{2} \chi}{3} \hat{q}^{3}} \\ 
&= \text{(phase)}\,  \hat{Z}\left[-\sqrt{2}\chi\left(m_{a}+\sqrt{2}m_{f}\right)^{2}\right]\hat{P}\left(4\chi m_{a} + 4\sqrt{2} \chi m_{f}\right)\hat{K}\left(-2\sqrt{2}\chi\right).
 \end{align}
Once plugged into Eq.~(\ref{eq:Lsimpphi}), the $\hat{Z}$ operator can be commuted to the left.  We now give the final result:
\begin{align}
\hat{L}(\chi, \sigma, \mathbf{m})&=\hat{Z}\left(\sqrt{2}m_{a}\right)\hat{X}(\kappa)\hat{R}\left( - \frac{\pi}{2} \right)\hat{P}\left[4\sigma+4 \chi \left(m_{a} +\sqrt{2} m_{f}\right)\right] \hat{K}\left(-2\sqrt{2}\chi\right)
,
\end{align}
 where we have now neglected (rather than merely omitted) the overall phase, and where
 \begin{align}
	\kappa = -2 m_{e}   \sqrt{1 + \sigma^2}  -2 \sigma(\sqrt{2}m_{a} + m_{f})-\sqrt{2}\chi(m_{a}+\sqrt{2}m_{f})^{2}
\end{align}
 from Eq.~\eqref{eq:kappadef}. This result is reported in Eq.~(\ref{eq:Kgate}).
\end{widetext}

\end{document}